\pdfoutput=1
\documentclass[aps,showpacs,preprintnumbers,amsmath,amssymb,twocolumn,superscriptaddress,prl]{revtex4-1}
\usepackage{txfonts}
\usepackage{amsmath}
\usepackage{float}
\usepackage{graphicx}
\usepackage{dcolumn}
\usepackage{bm}
\usepackage{color}
\usepackage{subfigure}  
\def\comment#1{}

\def\slashchar#1{\setbox0=\hbox{$#1$}           
	\dimen0=\wd0                                 
	\setbox1=\hbox{/} \dimen1=\wd1               
	\ifdim\dimen0>\dimen1                        
	\rlap{\hbox to \dimen0{\hfil/\hfil}}      
	#1                                        
	\else                                        
	\rlap{\hbox to \dimen1{\hfil$#1$\hfil}}   
	/                                         
	\fi}                                         %

\begin{document}

\title{Carbon nanotube field emission enhancement factors: links between experiment, classical theory and DFT-based induced-charge theory}

\author{Caio P. de Castro}
\email{caioporto@ufba.br}
\address{Instituto de F\'{\i}sica, Universidade Federal da Bahia,
   Campus Universit\'{a}rio da Federa\c c\~ao,
   Rua Bar\~{a}o de Jeremoabo s/n,
40170-115, Salvador, BA, Brazil}

\author{Thiago A. de Assis}
\email{thiagoaa@ufba.br}
\address{Instituto de F\'{\i}sica, Universidade Federal da Bahia,
   Campus Universit\'{a}rio da Federa\c c\~ao,
   Rua Bar\~{a}o de Jeremoabo s/n,
40170-115, Salvador, BA, Brazil}
\address{Centro Interdisciplinar em Energia e Ambiente, Universidade Federal da Bahia, Campus Universit\'{a}rio da Federa\c{c}\~{a}o, 40170-115 Salvador, BA, Brazil}

\author{Roberto Rivelino}
\email{rivelino@ufba.br}
\address{Instituto de F\'{\i}sica, Universidade Federal da Bahia,
   Campus Universit\'{a}rio da Federa\c c\~ao,
   Rua Bar\~{a}o de Jeremoabo s/n,
40170-115, Salvador, BA, Brazil}

\author{Fernando de B. Mota}
\email{fbmota@ufba.br}
\address{Instituto de F\'{\i}sica, Universidade Federal da Bahia,
   Campus Universit\'{a}rio da Federa\c c\~ao,
   Rua Bar\~{a}o de Jeremoabo s/n,
40170-115, Salvador, BA, Brazil}

\author{Caio M. C. de Castilho}
\email{caio@ufba.br}
\address{Instituto de F\'{\i}sica, Universidade Federal da Bahia,
   Campus Universit\'{a}rio da Federa\c c\~ao,
   Rua Bar\~{a}o de Jeremoabo s/n,
40170-115, Salvador, BA, Brazil}
\address{Centro Interdisciplinar em Energia e Ambiente, Universidade Federal da Bahia, Campus Universit\'{a}rio da Federa\c{c}\~{a}o, 40170-115 Salvador, BA, Brazil}
\address{Instituto Nacional de Ci\^{e}ncia e Tecnologia em Energia e Ambiente - INCTE\&A, Universidade Federal da Bahia,
   Campus Universit\'{a}rio da Federa\c c\~ao,
   Rua Bar\~{a}o de Jeremoabo s/n,
40170-280, Salvador, BA, Brazil}
\author{Richard G. Forbes}
\email{r.forbes@trinity.cantab.net}
\address{Advanced Technology Institute \& Department of Electrical and Electronic Engineering, University of Surrey, Guildford, Surrey GU2 7XH, UK}

\begin{abstract}
Experimental Fowler-Nordheim plots taken from orthodoxly behaving carbon nanotube (CNT) field electron emitters are known to be linear. This shows that, for such emitters, there exists a characteristic field enhancement factor (FEF) that is constant for a range of applied voltages and  applied macroscopic fields $F_\text{M}$. A constant FEF of this kind can be evaluated for classical CNT emitter models by finite-element and other methods, but (apparently contrary to experiment) several past quantum-mechanical (QM) CNT calculations find FEF-values that vary with $F_\text{M}$. A common feature of most such calculations is that they focus only on deriving the CNT real-charge distributions. Here we report on calculations that use density functional theory (DFT) to derive real-charge distributions, and then use these to generate the related induced-charge distributions and related fields and FEFs. We have analysed three carbon nanostructures involving CNT-like nanoprotrusions of various lengths, and have also simulated geometrically equivalent classical emitter models, using finite-element methods. We find that when the DFT-generated local induced FEFs (LIFEFs) are used, the resulting values are effectively independent of macroscopic field, and behave in the same qualitative manner as the classical FEF-values. Further, there is fair to good quantitative agreement between a characteristic FEF determined classically and the equivalent characteristic LIFEF generated via DFT approaches. Although many issues of detail remain to be explored, this appears to be a significant step forwards in linking classical and QM theories of CNT electrostatics. It also shows clearly that, for ideal CNTs, the known experimental constancy of the FEF value for a range of macroscopic fields can also be found in appropriately developed QM theory.
\end{abstract}

\maketitle

In discussions of field electron emission (FE) from a post-like carbon nanotube (CNT), field enhancement factors (FEFs) are important characterization parameters. A characteristic FEF $\gamma_\text{C}$ is defined as $\gamma_\text{C}\equiv F_\text{C}/F_\text{M}$, where $F_\text{M}$ is an applied macroscopic field, and $F_\text{C}$ is a particular local field, at a point C ``at the emitter surface''. The field $F_\text{C}$ specifies an electron tunneling barrier assumed to be characteristic of FE from the CNT.
Note that in this Letter ``local'' refers to a value at some particular point in three-dimensional space. Charges, charge-densities and electrostatic potentials have their conventional electrostatic meanings and signs (total charge near the field electron emitter apex is negative), but the symbol $F$ denotes a quantity (usually positive in FE) that is the negative of a conventional electrostatic field. We think this hybrid convention causes least confusion in the context of existing FE literature.

Careful experiments on emission from single CNTs (e.g., \cite{BonardPRL2002, Dean00, Xu05}) yield straight Fowler-Nordheim plots. The plots concerned all pass the orthodoxy test \cite{Forbes13}; hence, orthodox FN-plot analysis yields experimental estimates of $\gamma_\text{C}$ that are well-defined constants for the CNT under analysis. This is not necessarily expected for a CNT that has a small apex radius (less than about 10-20 nm) or for very long CNTs, so it is of interest that the measured Bonard CNT \cite{BonardPRL2002} had radius 7.5 nm and length 1.4 $\mu$m.
At a basic level of investigation, the observed apparent constancy of $\gamma_\text{C}$ (at least over a significant range of measured voltage and current) suggests the following. That, in theoretical modelling of CNTs, it ought to be possible to pick a definition of ``local field'' and a location ``C'' such that $F_\text{C}$$\propto$$F_\text{M}$. In order to ensure that no patch fields exist, classical-conductor models almost always make the additional simplifying assumption that the work-function is uniform across the conductor surface. In this case, classical electrostatics guarantees that $F_\text{C}$$\propto$$F_\text{M}$ at all points on the surface and at all points in space above it. Hence, in classical-conductor CNT models such as the ``hemisphere-on-a-cylindrical-post (HCP)'' model as used in parallel-plane geometry, it is usual to take point C at the post apex.

It is appealing to model a CNT as a post-like classical conductor. Fully reliable values of the local apex field $F_\text{a}$ and related FEF $\gamma_\text{a}$ can now be calculated by finite-element methods (e.g., \cite{Thiago}), and good approximate values by various other methods (e.g., \cite{Forbes03, Pogorelov10, Harris2015AIP}). However, few if any well-defined experiments exist with the same geometry as the usual classical models, so it is not currently possible to make reliable precise comparisons between equivalent experimental and theoretical FEF values.
More generally, there is no obvious a-priori reason to expect that a classical HCP or similar model should be a good model for a CNT emitter, since one might reasonably expect atomic-level or sub-atomic level considerations to be significant. Thus, the issue arises of how the apex field in a classical-conductor model relates to local fields in more realistic CNT models, such as those provided by the density functional theory (DFT) analysis of charged carbon nanostructures. This issue is the focus of this Letter.

Previous workers using atomistic models have identified the characteristic field $F_\text{C}$ in different ways. Thus, Mayer \cite{Mayer07}, using an essentially classical atomic-level methodology, identified $F_\text{C}$ as the effective field acting on the topmost carbon atom, in order to polarize it and give it a dipole moment. (In effect, Mayer's $F_\text{C}$ is the field acting at the nucleus of this topmost atom, due to all the other atoms.) As pointed out in \cite{Peng08}, this field is expected to be significantly less than the field outside the CNT.
Earlier, Buldum and Lu \cite{Buldum03} had used a quantum-mechanical (QM) self-consistent-field pseudo-potential methodology, and took $F_\text{C}$ as the field at the point where the electron emerged from the tunneling barrier. This approach leads to a FEF that varies with the macroscopic field $F_\text{M}$. The Sun Yat-Sen University group used a hybrid classical/QM method (see \cite{Zheng04} for details), and took $F_\text{C}$ to be the averaged field over a range of about 0.35 nm,  above the  emitter apex. This approach also leads to a FEF that varies with $F_\text{M}$. In a later (fully QM) version of this approach \cite{Peng08}, they took $F_\text{C}$ as defined at a fixed point above the emitter apex, but again found $F_\text{M}$-dependence in the FEF, and also numerical values that were significantly less than those predicted by an equivalent classical model. Amongst other factors, they suggested that these things might be due to field penetration and patch-field effects arising from the detailed electronic structure of the CNT.

The QM approaches just discussed all used calculated (but in fact approximate) real-charge distributions. However, it has been argued by one of us \cite{Forbes99} that -- in order to make good comparisons with classical models -- it is necessary to use induced-charge-density distributions $\rho_\text{i}(F_\text{M}, \textit{\textbf{R}})$, where $\textit{\textbf{R}}$ is an appropriate position vector. These are obtained by taking the difference between the real-charge-density distribution $\rho_\text{r}(F_\text{M}, \textit{\textbf{R}})$ in the presence of the macroscopic field, and the real-charge-density distribution $\rho_\text{r}(0, \textit{\textbf{R}})$ for zero macroscopic field, i.e., we set
\begin{equation}
\rho_\text{i}(F_\text{M}, \textit{\textbf{R}}) \equiv \rho_\text{r}(F_\text{M}, \textit{\textbf{R}}) - \rho_\text{r}(0, \textit{\textbf{R}}).
\end{equation}
In past work on planar metal surfaces, this induced charge distribution has been modeled by placing point charges and dipoles at the positions of the nuclei of surface atoms (e.g., \cite{Forbes80}). With the field present, the polarized dipoles cause the emitter's electrical surface (``the apparent zero-plane for classical electrostatic potential'') to be positioned outside the plane of the surface-atom nuclei by the repulsion distance $d_\text{rep}$ (which is typically about an atomic radius, as defined by half the nearest-neighbor distance in the crystal lattice). This ``electrical surface'' is at the electrical centroid of the (planar-surface) induced-charge distribution \cite{Forbes99}. For consistency in relation to electron potential energies, the surface of a classical-conductor model has to be made coincident with the emitter's electrical surface.
\begin{figure}[H]
\hspace{-0.7cm}
    \includegraphics[width=0.6\textwidth]{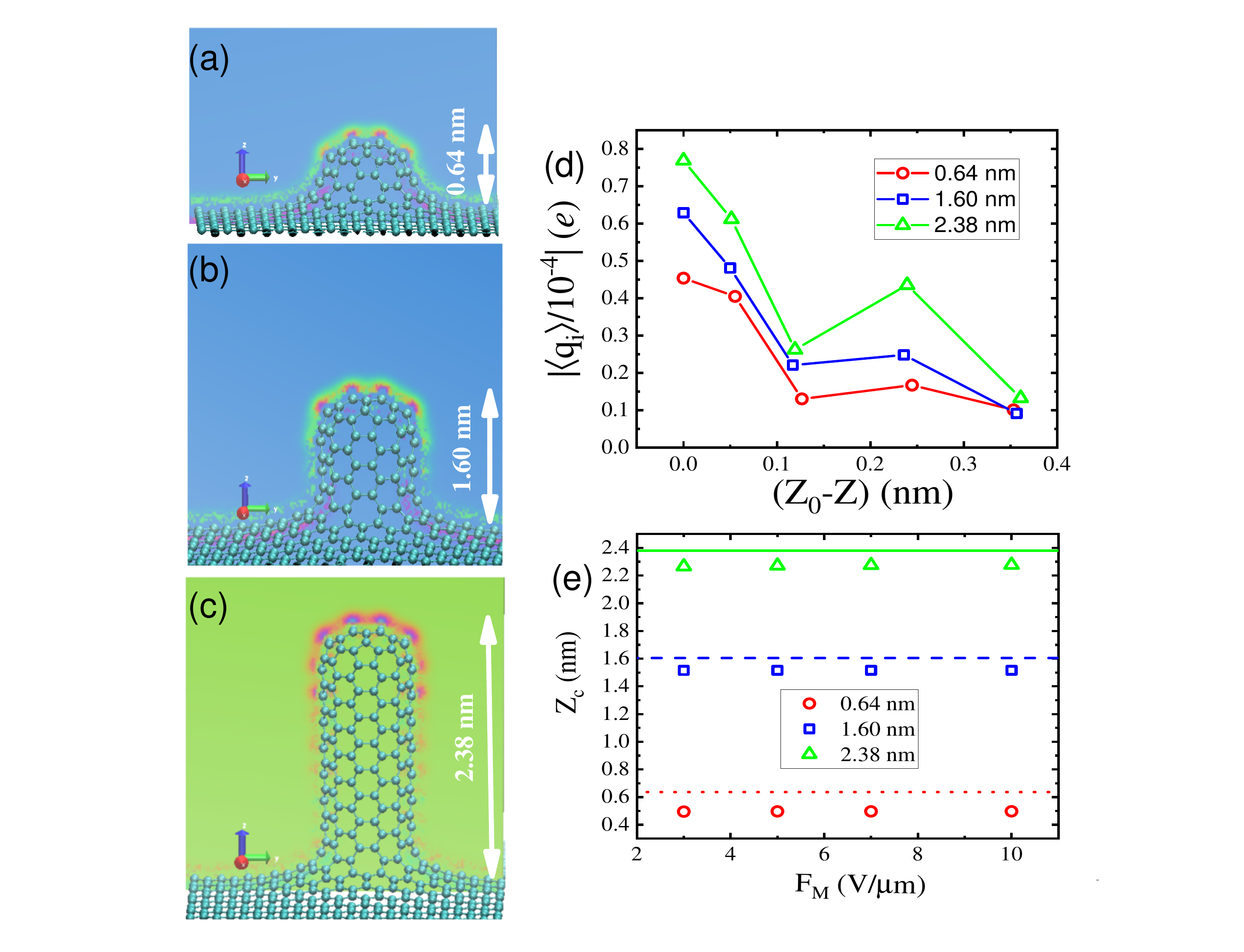}
	\caption{Seamless 3D graphene/nanotube hybrids using a capped (6,6) SWCNT. The protrusions have radius approximately 0.4 nm and heights of (a) 0.64 nm, (b) 1.12 nm, and (c) 2.38 nm. The background color maps show the magnitudes of the local induced charge-densities when a macroscopic field $F_\text{M}$ = 10 V/$\mu$m is applied. The images were generated by using Visual Molecular Dynamics (VMD) \cite{VMD}. (d) Magnitudes of the average induced charge, $\langle q_\text{i}\rangle$, obtained by integrating $\rho_\text{i}(F_\text{M},\textit{\textbf{R}})$ for the apex ring as a function of the position of each ring relative to the apex plane at $Z_0$. (e) Position $Z_\text{c}$ of the centroid of the induced charge, as a function of field $F_\text{M}$. The horizontal lines correspond to $Z_0$ for each structure.} \label{Systems}
\end{figure}
Lepetit \cite{Lepetit17} has recently carried out DFT calculations relating to field electron and ion emission on pyramidal metal surfaces. He also concludes that the correct place to define a characteristic local field $F_\text{C}$ is at the centroid of the induced-charge distribution (he calls this the ``induced-charge barycenter''). For the apex atoms of his double-pyramid, he finds repulsion distances similar to that derived using simple classical methods as just described (but different on the positive and negative sides).

This Letter reports an approach, based on first-principles calculations, in which we apply to a carbon nanostructure generally similar thinking about the use of the induced charge distribution. For this purpose we use a seamless three-dimensional (3D) graphene-CNT hybrid structure, as illustrated in Fig. \ref{Systems}.  This is intended as a model for a type of practical nanostructure, the so-called vertically aligned CNT on graphene (VAGCNT) introduced by Talapatra et al. \cite{Talapatra06}. We use ``Z" to denote distance measured along the VAGCNT vertical axis, and measure  $\textit{\textbf{R}}$ and $Z$-component from the point where the VAGCNT vertical axis intersects the VAGCNT base-plane.

\begin{figure}[H]
\hspace{-0.6cm}
    \includegraphics[width=0.60\textwidth]{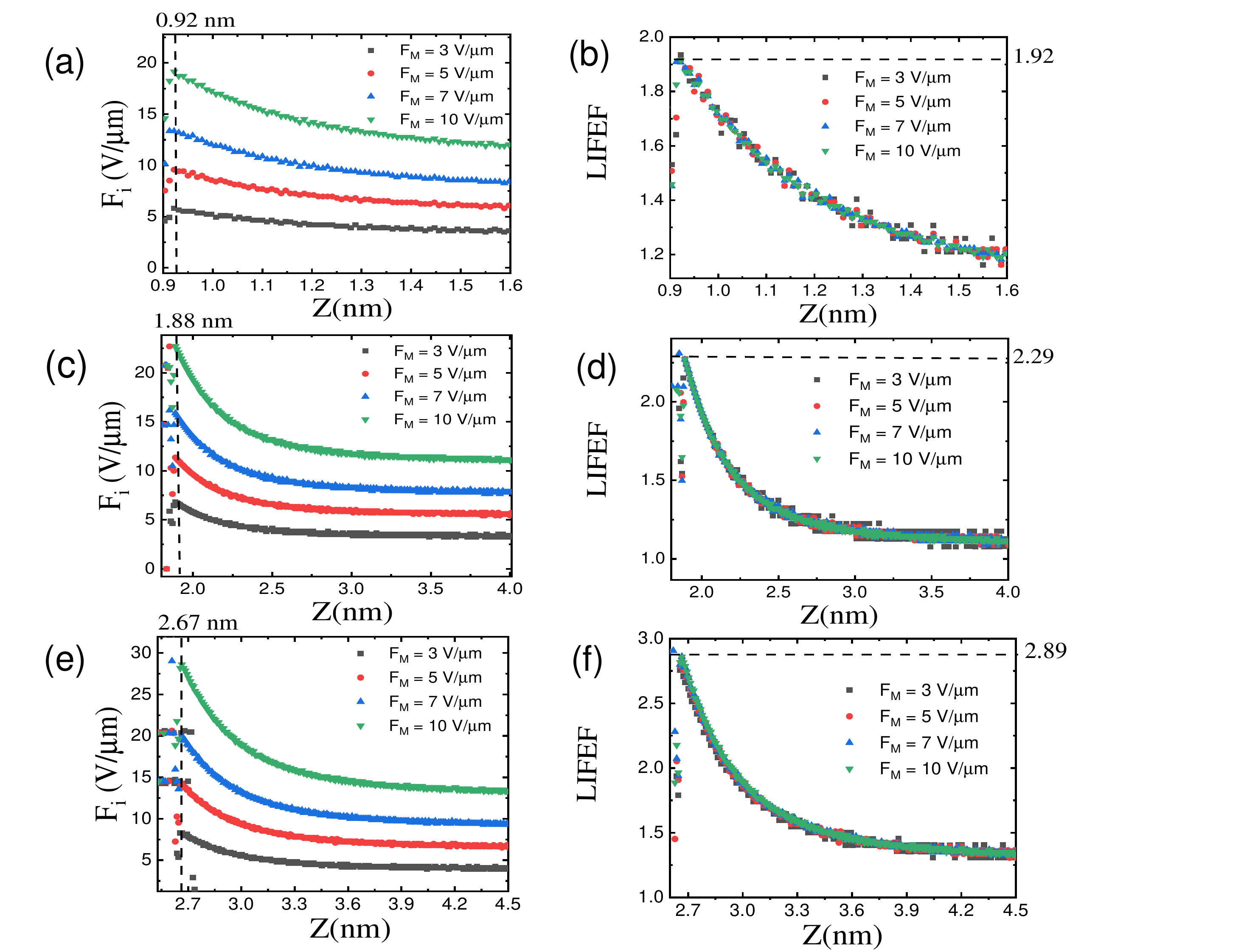}
	\caption{Magnitudes of local induced electrostatic field (LIEF) [(a),(c),(e)] and local induced field enhancement factor (LIFEF) [(b),(d),(f)] for $F_\text{M}$-values shown, for the three structures shown in Fig.\ref{Systems} (a) to (c). The positions of the maximum LIEF magnitudes (vertical dashed lines) and the values of the LIFEF (horizontal dashed lines) are indicated.}	\label{Results}
\end{figure}
In practical applications, this nanostructure sits on a metal base-plate. Thus, in principle -- in order to create a DFT model of this nanostructure in practical operation -- it would be necessary to include in the model the electrical effects of an image of the nanostructure in the base-plate. Alternatively, a more general approach of interest for FE experiment would require a protrusion subjected to two electrodes with different electrostatic potentials. This could be modeled self-consistently via, for example, a non-equilibrium Green's function (NEGF) formalism combined with DFT calculations \cite{Transiesta}. However, our real interest is in what would happen at the tip of a long CNT, for which image potentials would have negligible effect. Thus, we consider it a better first approximation not to include an image of the VAGCNT nanostructure in our calculations. What we shall be able to show is that if one considers the induced-charge-density distribution, rather than the real-charge-density distribution, then one can retrieve the classical electrostatic behavior of having a characteristic FEF that is independent of the applied macroscopic field.

Using the SIESTA DFT code \cite{SIESTA}, we deal with the VAGCNT structures as follows. First, we fully optimize the chosen geometry. In this process, atoms are allowed to relax until atomic forces decrease below 0.5 eV/nm. Systems of different lengths are defined within rectangular boxes with sizes $3.44 \times 2.98 \times 3.00$ nm$^{3}$, $3.43 \times 2.97 \times 5.00$ nm$^{3}$, and $3.43 \times 2.97 \times 6.00$ nm$^{3}$, respectively. The calculations assume a uniform applied ``macroscopic" electrostatic field (EF) oriented along the VAGCNT axis, with magnitude varying between 3 and 10 V/${\mathrm{\mu}}$m. The carbon atoms [of the unit cells] are described in terms of norm-conserving pseudopotentials and double-$\zeta$ basis sets, including polarization functions, with energy cutoff of 300 Ry, sampling the $\Gamma$ point of the Brillouin-zone. To solve the Kohn-Sham equations, we adopt the PBE exchange-correlation potential \cite{PBE}, which has performed well for carbon structures in applied fields \cite{Renato2017}.

Figure \ref{Systems} (a)-(c) shows the optimized 3D VAGCNTs. These behave as carbon nanoprotrusions (CNPs) formed by capped (6,6) single-walled CNTs (SWCNTs), with different heights relative to the 2D graphene sheet. The DFT calculations yield ``real atomic-level" values of local charge density $\rho_\text{r}(F_\text{M}, \textit{\textbf{R}})$ and local electrostatic potential  ${\mit{\Phi}}_\text{r}(F_\text{M}, \textit{\textbf{R}})$. From these, the norm $F_\text{r}(F_\text{M}, \textit{\textbf{R}})$ of the local ``real" electrostatic field can be derived. All these things depend on the macroscopic field $F_\text{M}$.
One can also use these calculated quantities to determine the related $induced$ quantities. Thus, the local induced charge-density $\rho_\text{i}(F_\text{M}, \textit{\textbf{R}})$ is given by eq. (1). Color maps in the plane that bisects the nanostructure can be used to display $\rho_\text{i}(F_\text{M}, \textit{\textbf{R}})$, as in Fig. \ref{Systems}. Our results confirm that, in line with classical expectation, the induced charges are concentrated at the CNP tips, and the overall trend is induced-charge decay from the CNP cap to its body, as observed in Fig. \ref{Systems}(d).

For the apex carbon-atom ring (parallel to the graphene plane), and the next four rings down the CNP, we  estimate the average induced charge per atom $\langle q_{i}\rangle$ for each ring. As expected, the charge magnitudes increase as the CNP height increases. The charge magnitudes decrease as we go down the CNP, except that (for all three CNPs) there is an induced-charge-magnitude increase from the third to the fourth ring. The fourth ring marks the start of the CNP cylindrical body, and the effect seems to relate to the change from roughly hemispherical to roughly cylindrical shape.
We can define an electrical centroid for the total induced charge on the apex ring and next four rings, taken together. As shown in Fig. \ref{Systems}(e), this centroid is just inside the CNT cap (just inside the apex ring), and its distance from the plane of the apex ring is effectively independent of macroscopic field, over the range of fields considered.
\begin{figure}[H]
\hspace{-0.2cm}
    \includegraphics[width=0.53\textwidth]{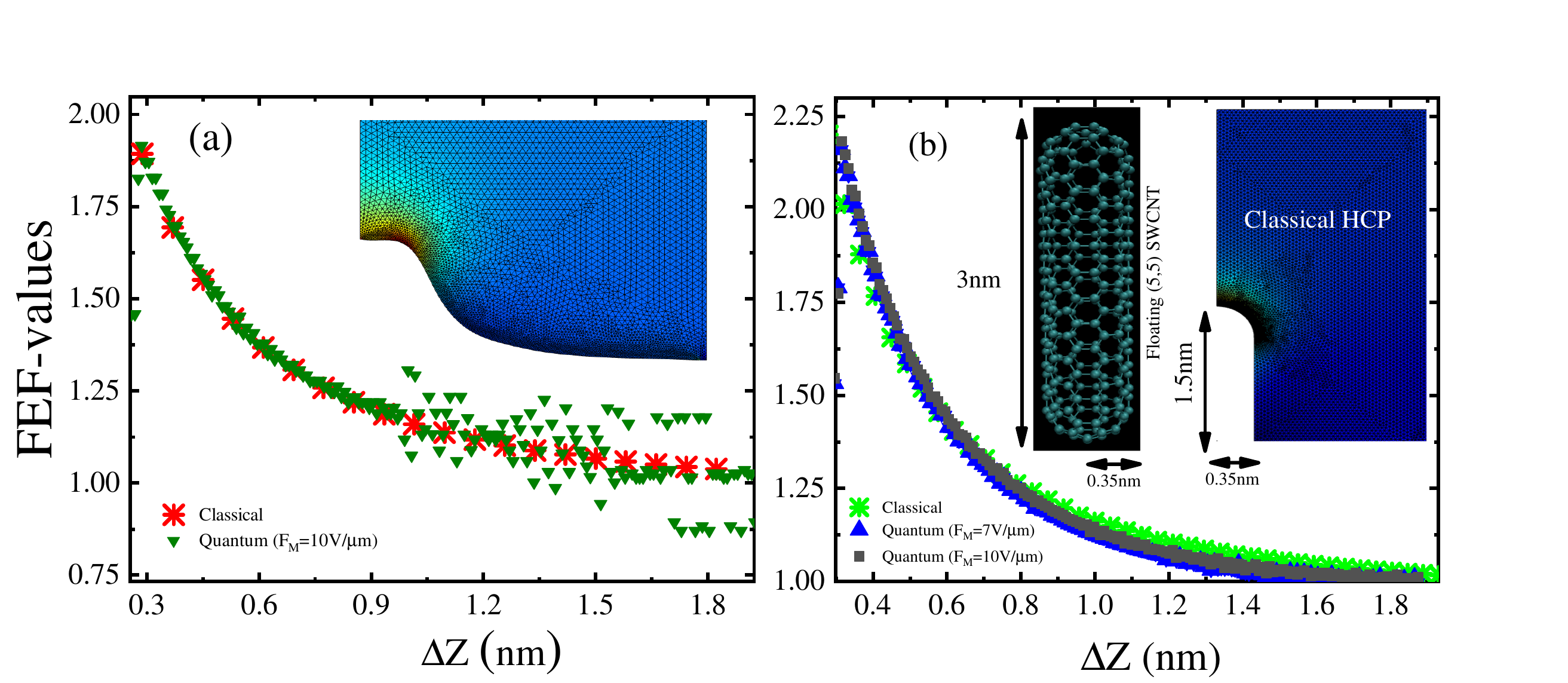}
	\caption{Local FEFs calculated classically, and local induced FEFs calculated by DFT, as functions (respectively) of ${\Delta}Z=Z-Z_\text{a}$ and ${\Delta}Z=Z-Z_\text{0}$ (see text for definitions), for (a) the smallest VAGCNT and (b) a floating (5,5) SWCNT. The insets also show half the classical protrusion, the mesh used, and a color map of the distribution of classical local FEF values; red (blue) pattern indicates higher (lower) FEF-values.}	 \label{Comparison}
\end{figure}
By analogy with eq.(1), we can define (the negative $F_\text{i}(F_\text{M}, \textit{\textbf{R}})$ of) a local induced electrostatic field (LIEF) by
\begin{equation}
F_\text{i}(F_\text{M}, \textit{\textbf{R}}) \equiv F_\text{r}(F_\text{M}, \textit{\textbf{R}}) - F_\text{r}(0, \textit{\textbf{R}}).
\end{equation}
For the structures shown in Fig.\ref{Systems}, for distances measured outwards from the plane (at $Z$ = $Z_0$) of the apex carbon-atom ring, Fig.\ref{Results} shows how the LIEF varies with distance $Z$ along the CNP axis. Also shown is the variation (along the axis) of the related local induced FEF (LIFEF), $\gamma_\text{i} \equiv F_\text{i}(F_\text{M}, \textit{\textbf{R}})/F_\text{M}$.
These DFT-derived profiles of the LIEF and LIFEF exhibit interesting and important features. The three VAGCNT models have different CNP lengths, but in each case the LIEF magnitude has a maximum value at about ($Z-Z_0$) = 0.28 nm. Further, for a given model, this maximum value does not depend on $F_\text{M}$. For ($Z-Z_0$) $>$ 0.28 nm, the LIEF magnitude decreases monotonically with $Z$ and the LIFEF tends asymptotically to unity, as expected classically. Further, for a given model, the nature of the dependence of the LIFEF on $Z$ is the same for all $F_\text{M}$-values, as expected in classical models. Thus, for a given CNP model, the different LIEF curves in the left-hand panels collapse into a single LIFEF curve in the right-hand panel. As far as we know, this is the first explicit demonstration that this collapse behavior can be found in first-principles electronic structure calculations on carbon nanostructures, when induced-charge (rather than real-charge) distributions are investigated.

In a CNT classical-conductor model, such as the HCP model, the local field has a maximum in the conductor surface, at the model apex. The characteristic local field and FEF are normally defined at this point. This supports the idea that, with a DFT model of a ``cylindrical'' CNP, we might usefully define characteristic LIEF and LIFEF values by taking them at the position of ``on-axis maximum field-magnitude". That is, we can usefully define a characteristic LIFEF by
\begin{equation}
\gamma_\text{i,C} \equiv \frac{\max{\left\{F_\text{i}\right\}}}{F_\text{M}}.
\label{CFEF}
\end{equation}
As shown in Fig. \ref{Results}, $\gamma_\text{i,C}$ increases from 1.92 to 2.89 as the CNP length increases from 0.64 nm to 2.38 nm, thus exhibiting the trend expected classically.
As a check, we have investigated the effects of simulation-box size. For the shortest structure, an increase of 40\% in the graphene base-area causes an increase of $\sim$ 5\% in $\gamma_\text{i,C}$. This shows that the systems considered here can be seen as arrays of VAGCNTs in which small depolarization effects operate.

Next, we used the shortest VAGCNT to compare DFT-calculated LIFEF-values with values calculated for a near-equivalent classical model. A continuous profile was drawn in a plane bisecting the VAGCNT vertically, by interpolating the  atomic positions, as shown in the inset of Fig. \ref{Comparison}(a). This profile was revolved by $2\pi$ around the symmetry axis, to generate a three dimensional body. In order to model the effect in such a fashion that the electrostatic potential immediately outside the surface of the base-plate is uniform, we have also considered the DFT-based electrostatic potential of a floating (5,5) SWCNT, as shown in the inset of Fig. \ref{Comparison}(b). We notice that the corresponding classical counterpart is an HCP model with post height equal to half the length of the floating nanotube, solved in a fully 3D domain.
For the classical models, Laplace's equation was then solved numerically, using finite element methods \cite{DeAssisNL,Thiago}, using a simulation box of size consistent with the VAGCNT and (5,5) SWCNT nanostructures used in the DFT calculations. Values of the resulting classical FEF $\gamma_\text{cl}$ are shown as a function of the variable $\Delta Z=Z-Z_\text{a}$, where $Z_\text{a}$ is the $Z$-coordinate of the apex of the classical profile. Also shown, but as a function of $\Delta Z = Z - Z_\text{0}$, are values of the DFT-calculated LIFEF $\gamma_\text{i,C}$.

As far the electrostatics of carbon nanostructures is concerned, the present results exhibit fair to good agreement between the DFT and classical calculation methodologies, when quantities related to the DFT-calculated $induced$-charge distributions are used. Strictly, in the classical modelling, the classical tip profile should be drawn slightly outside the nuclei of the apex carbon atoms, in order to be consistent with the idea that the electrical surface is repelled outwards by the polarization of the surface atoms, as discussed earlier. This would have the effect of moving the classical curves in Fig. \ref{Comparison} slightly to the left, but would not significantly affect the ``fair agreement" found. Since there are still some unresolved technical difficulties in deciding precisely how the electrical surface should be identified for a CNP, we have provisionally used the approximation of defining the classical tip profile by the positions of the surface atom nuclei. Even with this approximation, the results appear to suggest that the HCP model is a satisfactory model for the classical electrostatics of a CNT or CNP, and thus provide justification for using this classical model when discussing the technological development and applications of carbon-based field emitters.

Clearly, the results presented here are only a first step towards a reliable QM theory of CNT current-voltage characteristics. There remain unexplored detailed issues, including those ones concerning the relationships between (a) ``induced fields'' and ``real fields'', and (b) the shapes of ``induced-potential tunnelling barriers'' and ``real-potential tunneling barriers'' (presumably the latter one will determine actual electron transmission probabilities). Nevertheless, we think it important that, for carbon nanoprotrusions, by using induced-charge distributions, we have made a link between characteristic field enhancement factors calculated classically and those calculated using density functional theory. It also appears to be the first time that the experimentally observed constancy of characteristic FEF for an orthodoxly behaving CNT, for a significant range of applied macroscopic fields, has been clearly seen in a related quantum-mechanical calculation.

This study was partially financed by Coordena\c{c}\~{a}o de Aperfei\c{c}oamento de Pessoal de N\'{i}vel Superior - Brasil
(CAPES) - Finance Code 001, and by the  CNPq (Brazilian Agency). The authors would also like to thank the Royal Society of London for financial support via a Newton Mobility Grant, Ref: NI160031.


\begin{thebibliography}{22}%
\makeatletter
\providecommand \@ifxundefined [1]{%
 \@ifx{#1\undefined}
}%
\providecommand \@ifnum [1]{%
 \ifnum #1\expandafter \@firstoftwo
 \else \expandafter \@secondoftwo
 \fi
}%
\providecommand \@ifx [1]{%
 \ifx #1\expandafter \@firstoftwo
 \else \expandafter \@secondoftwo
 \fi
}%
\providecommand \natexlab [1]{#1}%
\providecommand \enquote  [1]{``#1''}%
\providecommand \bibnamefont  [1]{#1}%
\providecommand \bibfnamefont [1]{#1}%
\providecommand \citenamefont [1]{#1}%
\providecommand \href@noop [0]{\@secondoftwo}%
\providecommand \href [0]{\begingroup \@sanitize@url \@href}%
\providecommand \@href[1]{\@@startlink{#1}\@@href}%
\providecommand \@@href[1]{\endgroup#1\@@endlink}%
\providecommand \@sanitize@url [0]{\catcode `\\12\catcode `\$12\catcode
  `\&12\catcode `\#12\catcode `\^12\catcode `\_12\catcode `\%12\relax}%
\providecommand \@@startlink[1]{}%
\providecommand \@@endlink[0]{}%
\providecommand \url  [0]{\begingroup\@sanitize@url \@url }%
\providecommand \@url [1]{\endgroup\@href {#1}{\urlprefix }}%
\providecommand \urlprefix  [0]{URL }%
\providecommand \Eprint [0]{\href }%
\providecommand \doibase [0]{http://dx.doi.org/}%
\providecommand \selectlanguage [0]{\@gobble}%
\providecommand \bibinfo  [0]{\@secondoftwo}%
\providecommand \bibfield  [0]{\@secondoftwo}%
\providecommand \translation [1]{[#1]}%
\providecommand \BibitemOpen [0]{}%
\providecommand \bibitemStop [0]{}%
\providecommand \bibitemNoStop [0]{.\EOS\space}%
\providecommand \EOS [0]{\spacefactor3000\relax}%
\providecommand \BibitemShut  [1]{\csname bibitem#1\endcsname}%
\let\auto@bib@innerbib\@empty
\bibitem [{\citenamefont {Bonard}\ \emph {et~al.}(2002)\citenamefont {Bonard},
  \citenamefont {Dean}, \citenamefont {Coll},\ and\ \citenamefont
  {Klinke}}]{BonardPRL2002}%
  \BibitemOpen
  \bibfield  {author} {\bibinfo {author} {\bibfnamefont {J.-M.}\ \bibnamefont
  {Bonard}}, \bibinfo {author} {\bibfnamefont {K.~A.}\ \bibnamefont {Dean}},
  \bibinfo {author} {\bibfnamefont {B.~F.}\ \bibnamefont {Coll}}, \ and\
  \bibinfo {author} {\bibfnamefont {C.}~\bibnamefont {Klinke}},\ }\href
  {\doibase 10.1103/PhysRevLett.89.197602} {\bibfield  {journal} {\bibinfo
  {journal} {Phys. Rev. Lett.}\ }\textbf {\bibinfo {volume} {89}},\ \bibinfo
  {pages} {197602} (\bibinfo {year} {2002})}\BibitemShut {NoStop}%
\bibitem [{\citenamefont {Dean}\ and\ \citenamefont
  {Chalamala}(2000)}]{Dean00}%
  \BibitemOpen
  \bibfield  {author} {\bibinfo {author} {\bibfnamefont {K.~A.}\ \bibnamefont
  {Dean}}\ and\ \bibinfo {author} {\bibfnamefont {B.~R.}\ \bibnamefont
  {Chalamala}},\ }\href {\doibase 10.1063/1.125758} {\bibfield  {journal}
  {\bibinfo  {journal} {Applied Physics Letters}\ }\textbf {\bibinfo {volume}
  {76}},\ \bibinfo {pages} {375} (\bibinfo {year} {2000})}\BibitemShut
  {NoStop}%
\bibitem [{\citenamefont {Xu}\ \emph {et~al.}(2005)\citenamefont {Xu},
  \citenamefont {Bai}, \citenamefont {Wang},\ and\ \citenamefont
  {Wang}}]{Xu05}%
  \BibitemOpen
  \bibfield  {author} {\bibinfo {author} {\bibfnamefont {Z.}~\bibnamefont
  {Xu}}, \bibinfo {author} {\bibfnamefont {X.~D.}\ \bibnamefont {Bai}},
  \bibinfo {author} {\bibfnamefont {E.~G.}\ \bibnamefont {Wang}}, \ and\
  \bibinfo {author} {\bibfnamefont {Z.~L.}\ \bibnamefont {Wang}},\ }\href
  {\doibase 10.1063/1.2103420} {\bibfield  {journal} {\bibinfo  {journal}
  {Applied Physics Letters}\ }\textbf {\bibinfo {volume} {87}},\ \bibinfo
  {pages} {163106} (\bibinfo {year} {2005})}\BibitemShut {NoStop}%
\bibitem [{\citenamefont {Forbes}(2013)}]{Forbes13}%
  \BibitemOpen
  \bibfield  {author} {\bibinfo {author} {\bibfnamefont {R.~G.}\ \bibnamefont
  {Forbes}},\ }\href@noop {} {\bibfield  {journal} {\bibinfo  {journal}
  {Proceedings of the Royal Society of London A: Mathematical, Physical and
  Engineering Sciences}\ }\textbf {\bibinfo {volume} {469}},\ \bibinfo {pages}
  {20130271} (\bibinfo {year} {2013})}\BibitemShut {NoStop}%
\bibitem [{\citenamefont {de~Assis}\ and\ \citenamefont
  {Dall'Agnol}(2018)}]{Thiago}%
  \BibitemOpen
  \bibfield  {author} {\bibinfo {author} {\bibfnamefont {T.~A.}\ \bibnamefont
  {de~Assis}}\ and\ \bibinfo {author} {\bibfnamefont {F.~F.}\ \bibnamefont
  {Dall'Agnol}},\ }\href@noop {} {\bibfield  {journal} {\bibinfo  {journal}
  {Journal of Physics: Condensed Matter}\ }\textbf {\bibinfo {volume} {30}},\
  \bibinfo {pages} {195301} (\bibinfo {year} {2018})}\BibitemShut {NoStop}%
\bibitem [{\citenamefont {Forbes}\ \emph {et~al.}(2003)\citenamefont {Forbes},
  \citenamefont {Edgcombe},\ and\ \citenamefont {Valdr\`{e}}}]{Forbes03}%
  \BibitemOpen
  \bibfield  {author} {\bibinfo {author} {\bibfnamefont {R.~G.}\ \bibnamefont
  {Forbes}}, \bibinfo {author} {\bibfnamefont {C.}~\bibnamefont {Edgcombe}}, \
  and\ \bibinfo {author} {\bibfnamefont {U.}~\bibnamefont {Valdr\`{e}}},\
  }\href {\doibase http://dx.doi.org/10.1016/S0304-3991(02)00297-8} {\bibfield
  {journal} {\bibinfo  {journal} {Ultramicroscopy}\ }\textbf {\bibinfo {volume}
  {95}},\ \bibinfo {pages} {57 } (\bibinfo {year} {2003})}\BibitemShut
  {NoStop}%
\bibitem [{\citenamefont {Pogorelov}\ \emph {et~al.}(2010)\citenamefont
  {Pogorelov}, \citenamefont {Chang}, \citenamefont {Zhbanov},\ and\
  \citenamefont {Lee}}]{Pogorelov10}%
  \BibitemOpen
  \bibfield  {author} {\bibinfo {author} {\bibfnamefont {E.~G.}\ \bibnamefont
  {Pogorelov}}, \bibinfo {author} {\bibfnamefont {Y.-C.}\ \bibnamefont
  {Chang}}, \bibinfo {author} {\bibfnamefont {A.~I.}\ \bibnamefont {Zhbanov}},
  \ and\ \bibinfo {author} {\bibfnamefont {Y.-G.}\ \bibnamefont {Lee}},\ }\href
  {\doibase 10.1063/1.3466992} {\bibfield  {journal} {\bibinfo  {journal}
  {Journal of Applied Physics}\ }\textbf {\bibinfo {volume} {108}},\ \bibinfo
  {pages} {044502} (\bibinfo {year} {2010})}\BibitemShut {NoStop}%
\bibitem [{\citenamefont {Harris}\ \emph {et~al.}(2015)\citenamefont {Harris},
  \citenamefont {Jensen},\ and\ \citenamefont {Shiffler}}]{Harris2015AIP}%
  \BibitemOpen
  \bibfield  {author} {\bibinfo {author} {\bibfnamefont {J.~R.}\ \bibnamefont
  {Harris}}, \bibinfo {author} {\bibfnamefont {K.~L.}\ \bibnamefont {Jensen}},
  \ and\ \bibinfo {author} {\bibfnamefont {D.~A.}\ \bibnamefont {Shiffler}},\
  }\href {\doibase 10.1063/1.4929983} {\bibfield  {journal} {\bibinfo
  {journal} {AIP Advances}\ }\textbf {\bibinfo {volume} {5}},\ \bibinfo {pages}
  {087182} (\bibinfo {year} {2015})}\BibitemShut {NoStop}%
\bibitem [{\citenamefont {Mayer}(2007)}]{Mayer07}%
  \BibitemOpen
  \bibfield  {author} {\bibinfo {author} {\bibfnamefont {A.}~\bibnamefont
  {Mayer}},\ }\href {\doibase 10.1103/PhysRevB.75.045407} {\bibfield  {journal}
  {\bibinfo  {journal} {Phys. Rev. B}\ }\textbf {\bibinfo {volume} {75}},\
  \bibinfo {pages} {045407} (\bibinfo {year} {2007})}\BibitemShut {NoStop}%
\bibitem [{\citenamefont {Peng}\ \emph {et~al.}(2008)\citenamefont {Peng},
  \citenamefont {Li}, \citenamefont {He}, \citenamefont {Chen}, \citenamefont
  {Wang}, \citenamefont {Deng}, \citenamefont {Xu}, \citenamefont {Zheng},
  \citenamefont {Chen}, \citenamefont {Edgcombe},\ and\ \citenamefont
  {Forbes}}]{Peng08}%
  \BibitemOpen
  \bibfield  {author} {\bibinfo {author} {\bibfnamefont {J.}~\bibnamefont
  {Peng}}, \bibinfo {author} {\bibfnamefont {Z.}~\bibnamefont {Li}}, \bibinfo
  {author} {\bibfnamefont {C.}~\bibnamefont {He}}, \bibinfo {author}
  {\bibfnamefont {G.}~\bibnamefont {Chen}}, \bibinfo {author} {\bibfnamefont
  {W.}~\bibnamefont {Wang}}, \bibinfo {author} {\bibfnamefont {S.}~\bibnamefont
  {Deng}}, \bibinfo {author} {\bibfnamefont {N.}~\bibnamefont {Xu}}, \bibinfo
  {author} {\bibfnamefont {X.}~\bibnamefont {Zheng}}, \bibinfo {author}
  {\bibfnamefont {G.}~\bibnamefont {Chen}}, \bibinfo {author} {\bibfnamefont
  {C.~J.}\ \bibnamefont {Edgcombe}}, \ and\ \bibinfo {author} {\bibfnamefont
  {R.~G.}\ \bibnamefont {Forbes}},\ }\href {\doibase 10.1063/1.2946449}
  {\bibfield  {journal} {\bibinfo  {journal} {Journal of Applied Physics}\
  }\textbf {\bibinfo {volume} {104}},\ \bibinfo {pages} {014310} (\bibinfo
  {year} {2008})}\BibitemShut {NoStop}%
\bibitem [{\citenamefont {Buldum}\ and\ \citenamefont {Lu}(2003)}]{Buldum03}%
  \BibitemOpen
  \bibfield  {author} {\bibinfo {author} {\bibfnamefont {A.}~\bibnamefont
  {Buldum}}\ and\ \bibinfo {author} {\bibfnamefont {J.~P.}\ \bibnamefont
  {Lu}},\ }\href {\doibase 10.1103/PhysRevLett.91.236801} {\bibfield  {journal}
  {\bibinfo  {journal} {Phys. Rev. Lett.}\ }\textbf {\bibinfo {volume} {91}},\
  \bibinfo {pages} {236801} (\bibinfo {year} {2003})}\BibitemShut {NoStop}%
\bibitem [{\citenamefont {Zheng}\ \emph {et~al.}(2004)\citenamefont {Zheng},
  \citenamefont {Chen}, \citenamefont {Li}, \citenamefont {Deng},\ and\
  \citenamefont {Xu}}]{Zheng04}%
  \BibitemOpen
  \bibfield  {author} {\bibinfo {author} {\bibfnamefont {X.}~\bibnamefont
  {Zheng}}, \bibinfo {author} {\bibfnamefont {G.}~\bibnamefont {Chen}},
  \bibinfo {author} {\bibfnamefont {Z.}~\bibnamefont {Li}}, \bibinfo {author}
  {\bibfnamefont {S.}~\bibnamefont {Deng}}, \ and\ \bibinfo {author}
  {\bibfnamefont {N.}~\bibnamefont {Xu}},\ }\href {\doibase
  10.1103/PhysRevLett.92.106803} {\bibfield  {journal} {\bibinfo  {journal}
  {Phys. Rev. Lett.}\ }\textbf {\bibinfo {volume} {92}},\ \bibinfo {pages}
  {106803} (\bibinfo {year} {2004})}\BibitemShut {NoStop}%
\bibitem [{\citenamefont {Forbes}(1999)}]{Forbes99}%
  \BibitemOpen
  \bibfield  {author} {\bibinfo {author} {\bibfnamefont {R.~G.}\ \bibnamefont
  {Forbes}},\ }\href {\doibase https://doi.org/10.1016/S0304-3991(99)00098-4}
  {\bibfield  {journal} {\bibinfo  {journal} {Ultramicroscopy}\ }\textbf
  {\bibinfo {volume} {79}},\ \bibinfo {pages} {25 } (\bibinfo {year}
  {1999})}\BibitemShut {NoStop}%
\bibitem [{\citenamefont {Forbes}(1980)}]{Forbes80}%
  \BibitemOpen
  \bibfield  {author} {\bibinfo {author} {\bibfnamefont {R.~G.}\ \bibnamefont
  {Forbes}},\ }\href {\doibase 10.1063/1.91633} {\bibfield  {journal} {\bibinfo
   {journal} {Applied Physics Letters}\ }\textbf {\bibinfo {volume} {36}},\
  \bibinfo {pages} {739} (\bibinfo {year} {1980})}\BibitemShut {NoStop}%
\bibitem [{\citenamefont {Humphrey}\ \emph {et~al.}(1996)\citenamefont
  {Humphrey}, \citenamefont {Dalke},\ and\ \citenamefont {Schulten}}]{VMD}%
  \BibitemOpen
  \bibfield  {author} {\bibinfo {author} {\bibfnamefont {W.}~\bibnamefont
  {Humphrey}}, \bibinfo {author} {\bibfnamefont {A.}~\bibnamefont {Dalke}}, \
  and\ \bibinfo {author} {\bibfnamefont {K.}~\bibnamefont {Schulten}},\ }\href
  {\doibase https://doi.org/10.1016/0263-7855(96)00018-5} {\bibfield  {journal}
  {\bibinfo  {journal} {Journal of Molecular Graphics}\ }\textbf {\bibinfo
  {volume} {14}},\ \bibinfo {pages} {33 } (\bibinfo {year} {1996})}\BibitemShut
  {NoStop}%
\bibitem [{\citenamefont {Lepetit}\ \emph {et~al.}(2016)\citenamefont
  {Lepetit}, \citenamefont {Lemoine},\ and\ \citenamefont
  {M?rquez-Mijares}}]{Lepetit17}%
  \BibitemOpen
  \bibfield  {author} {\bibinfo {author} {\bibfnamefont {B.}~\bibnamefont
  {Lepetit}}, \bibinfo {author} {\bibfnamefont {D.}~\bibnamefont {Lemoine}}, \
  and\ \bibinfo {author} {\bibfnamefont {M.}~\bibnamefont
  {M?rquez-Mijares}},\ }\href {\doibase 10.1063/1.4961216} {\bibfield
  {journal} {\bibinfo  {journal} {Journal of Applied Physics}\ }\textbf
  {\bibinfo {volume} {120}},\ \bibinfo {pages} {085105} (\bibinfo {year}
  {2016})}\BibitemShut {NoStop}%
\bibitem [{\citenamefont {Talapatra}\ \emph {et~al.}(2006)\citenamefont
  {Talapatra}, \citenamefont {Kar}, \citenamefont {Pal}, \citenamefont
  {Vajtai}, \citenamefont {Ci}, \citenamefont {Victor}, \citenamefont
  {Shaijumon}, \citenamefont {Kaur}, \citenamefont {Nalamasu},\ and\
  \citenamefont {Ajayan}}]{Talapatra06}%
  \BibitemOpen
  \bibfield  {author} {\bibinfo {author} {\bibfnamefont {S.}~\bibnamefont
  {Talapatra}}, \bibinfo {author} {\bibfnamefont {S.}~\bibnamefont {Kar}},
  \bibinfo {author} {\bibfnamefont {S.~K.}\ \bibnamefont {Pal}}, \bibinfo
  {author} {\bibfnamefont {R.}~\bibnamefont {Vajtai}}, \bibinfo {author}
  {\bibfnamefont {L.}~\bibnamefont {Ci}}, \bibinfo {author} {\bibfnamefont
  {P.}~\bibnamefont {Victor}}, \bibinfo {author} {\bibfnamefont {M.~M.}\
  \bibnamefont {Shaijumon}}, \bibinfo {author} {\bibfnamefont {S.}~\bibnamefont
  {Kaur}}, \bibinfo {author} {\bibfnamefont {O.}~\bibnamefont {Nalamasu}}, \
  and\ \bibinfo {author} {\bibfnamefont {P.~M.}\ \bibnamefont {Ajayan}},\
  }\href {\doibase http://dx.doi.org/10.1038/nnano.2006.56} {\bibfield
  {journal} {\bibinfo  {journal} {Nat. Nanotechnol.}\ }\textbf {\bibinfo
  {volume} {1}},\ \bibinfo {pages} {112} (\bibinfo {year} {2006})}\BibitemShut
  {NoStop}%
\bibitem [{\citenamefont {Brandbyge}\ \emph {et~al.}(2002)\citenamefont
  {Brandbyge}, \citenamefont {Mozos}, \citenamefont {Ordej\'on}, \citenamefont
  {Taylor},\ and\ \citenamefont {Stokbro}}]{Transiesta}%
  \BibitemOpen
  \bibfield  {author} {\bibinfo {author} {\bibfnamefont {M.}~\bibnamefont
  {Brandbyge}}, \bibinfo {author} {\bibfnamefont {J.-L.}\ \bibnamefont
  {Mozos}}, \bibinfo {author} {\bibfnamefont {P.}~\bibnamefont {Ordej\'on}},
  \bibinfo {author} {\bibfnamefont {J.}~\bibnamefont {Taylor}}, \ and\ \bibinfo
  {author} {\bibfnamefont {K.}~\bibnamefont {Stokbro}},\ }\href {\doibase
  10.1103/PhysRevB.65.165401} {\bibfield  {journal} {\bibinfo  {journal} {Phys.
  Rev. B}\ }\textbf {\bibinfo {volume} {65}},\ \bibinfo {pages} {165401}
  (\bibinfo {year} {2002})}\BibitemShut {NoStop}%
\bibitem [{\citenamefont {Soler}\ \emph {et~al.}(2002)\citenamefont {Soler},
  \citenamefont {Artacho}, \citenamefont {Gale}, \citenamefont {Garc?a},
  \citenamefont {Junquera}, \citenamefont {Ordej?n},\ and\ \citenamefont
  {S?nchez-Portal}}]{SIESTA}%
  \BibitemOpen
  \bibfield  {author} {\bibinfo {author} {\bibfnamefont {J.~M.}\ \bibnamefont
  {Soler}}, \bibinfo {author} {\bibfnamefont {E.}~\bibnamefont {Artacho}},
  \bibinfo {author} {\bibfnamefont {J.~D.}\ \bibnamefont {Gale}}, \bibinfo
  {author} {\bibfnamefont {A.}~\bibnamefont {Garc?a}}, \bibinfo {author}
  {\bibfnamefont {J.}~\bibnamefont {Junquera}}, \bibinfo {author}
  {\bibfnamefont {P.}~\bibnamefont {Ordej?n}}, \ and\ \bibinfo {author}
  {\bibfnamefont {D.}~\bibnamefont {S?nchez-Portal}},\ }\href@noop {}
  {\bibfield  {journal} {\bibinfo  {journal} {Journal of Physics: Condensed
  Matter}\ }\textbf {\bibinfo {volume} {14}},\ \bibinfo {pages} {2745}
  (\bibinfo {year} {2002})}\BibitemShut {NoStop}%
\bibitem [{\citenamefont {Perdew}\ \emph {et~al.}(1996)\citenamefont {Perdew},
  \citenamefont {Burke},\ and\ \citenamefont {Ernzerhof}}]{PBE}%
  \BibitemOpen
  \bibfield  {author} {\bibinfo {author} {\bibfnamefont {J.~P.}\ \bibnamefont
  {Perdew}}, \bibinfo {author} {\bibfnamefont {K.}~\bibnamefont {Burke}}, \
  and\ \bibinfo {author} {\bibfnamefont {M.}~\bibnamefont {Ernzerhof}},\ }\href
  {\doibase 10.1103/PhysRevLett.77.3865} {\bibfield  {journal} {\bibinfo
  {journal} {Phys. Rev. Lett.}\ }\textbf {\bibinfo {volume} {77}},\ \bibinfo
  {pages} {3865} (\bibinfo {year} {1996})}\BibitemShut {NoStop}%
\bibitem [{\citenamefont {dos Santos}\ \emph {et~al.}(2017)\citenamefont {dos
  Santos}, \citenamefont {Mota}, \citenamefont {Rivelino},\ and\ \citenamefont
  {Gueorguiev}}]{Renato2017}%
  \BibitemOpen
  \bibfield  {author} {\bibinfo {author} {\bibfnamefont {R.~B.}\ \bibnamefont
  {dos Santos}}, \bibinfo {author} {\bibfnamefont {F.~B.}\ \bibnamefont
  {Mota}}, \bibinfo {author} {\bibfnamefont {R.}~\bibnamefont {Rivelino}}, \
  and\ \bibinfo {author} {\bibfnamefont {G.~K.}\ \bibnamefont {Gueorguiev}},\
  }\href {\doibase 10.1021/acs.jpcc.7b09447} {\bibfield  {journal} {\bibinfo
  {journal} {J. Phys. Chem. C}\ }\textbf {\bibinfo {volume} {121}},\ \bibinfo
  {pages} {26125} (\bibinfo {year} {2017})}\BibitemShut {NoStop}%
\bibitem [{\citenamefont {de~Assis}\ and\ \citenamefont
  {Dall'Agnol}(2016)}]{DeAssisNL}%
  \BibitemOpen
  \bibfield  {author} {\bibinfo {author} {\bibfnamefont {T.~A.}\ \bibnamefont
  {de~Assis}}\ and\ \bibinfo {author} {\bibfnamefont {F.~F.}\ \bibnamefont
  {Dall'Agnol}},\ }\href {http://stacks.iop.org/0957-4484/27/i=44/a=44LT01}
  {\bibfield  {journal} {\bibinfo  {journal} {Nanotechnology}\ }\textbf
  {\bibinfo {volume} {27}},\ \bibinfo {pages} {44LT01} (\bibinfo {year}
  {2016})}\BibitemShut {NoStop}%
\end{thebibliography}

%

\end{document}